\documentclass[twocolumn,showpacs,superscriptaddress,amsmath,amssymb]{revtex4-1}
\usepackage{epsfig}
\usepackage{dcolumn}
\usepackage{bm}
\usepackage{color}

\def\bd{\begin{document}} \def\ed{\end{document}}
\def\bmp{\begin{minipage}} \def\emp{\end{minipage}}
\def\bcc{\begin{center}} \def\ecc{\end{center}}     \def\npg{\newpage}
\def\beq{\begin{equation}} \def\eeq{\end{equation}} \def\hph{\hphantom}
\def\be{\begin{equation}} \def\ee{\end{equation}} \def\r#1{$^{[#1]}$}
\def\n{\noindent} \def\ni{\noindent} \def\pa{\parindent}
\def\hs{\hskip} \def\vs{\vskip} \def\hf{\hfill} \def\ej{\vfill\eject}
\def\cl{\centerline} \def\ob{\obeylines}  \def\ls{\leftskip}
\def\underbar#1{$\setbox0=\hbox{#1} \dp0=1.5pt \mathsurround=0pt
   \underline{\box0}$}   \def\ub{\underbar}    \def\ul{\underline}
\def\f{\left} \def\g{\right} \def\e{{\rm e}} \def\o{\over} \def\d{{\rm d}}
\def\vf{\varphi} \def\pl{\partial} \def\cov{{\rm cov}} \def\ch{{\rm ch}}
\def\la{\langle} \def\ra{\rangle} \def\EE{e$^+$e$^-$} \def\pt{p_{\rm t}}
\def\bitz{\begin{itemize}} \def\eitz{\end{itemize}}
\def\btbl{\begin{tabular}} \def\etbl{\end{tabular}}
\def\btbb{\begin{tabbing}} \def\etbb{\end{tabbing}}
\def\beqar{\begin{eqnarray}} \def\eeqar{\end{eqnarray}}
\def\\{\hfill\break} \def\dit{\item{-}} \def\i{\item}
\def\bbb{\begin{thebibliography}{9} \itemsep=-1mm}
\def\ebb{\end{thebibliography}} \def\bb{\bibitem}
\def\bpic{\begin{picture}(260,240)} \def\epic{\end{picture}}
\def\akgt{\cl{\bf ACKNOWLEDGMENTS}}
\def\fgn{\noindent{\bf\large\bf figure captions}}
\def\lan{\langle}
\def\ran{\rangle}
\def\p{\pi}
\def\ifmath#1{\relax\ifmmode #1\else $#1$\fi}%
\def\rc{\ifmath{{\mathrm{c}}}}
\def\cut{\ifmath{{\mathrm{cut}}}}
\def\rF{\ifmath{{\mathrm{F}}}}
\def\rK{\ifmath{{\mathrm{K}}}}
\def\rp{\ifmath{{\mathrm{p}}}}
\def\rt{\ifmath{{\mathrm{t}}}}
\def\LAB{\ifmath{{\mathrm{LAB}}}}
\def\cut{\ifmath{{\mathrm{cut}}}}
\def\beq{\begin{equation}}
\def\eeq{\end{equation}}

\newcommand{\cinst}[2]{$^{\mathrm{#1}}$~#2\par}
\newcommand{\crefi}[1]{$^{\mathrm{#1}}$}
\newcommand{\crefii}[2]{$^{\mathrm{#1,#2}}$}
\newcommand{\crefiii}[3]{$^{\mathrm{#1,#2,#3}}$}
\newcommand{\HRule}{\rule{0.5\linewidth}{0.5mm}}

\bd

\title{\boldmath Measurements of $h_c(^1P_1)$ in $\psi'$ Decays}

\author{
{\small M.~Ablikim$^{1}$, M.~N.~Achasov$^{5}$, L.~An$^{9}$,
Q.~An$^{31}$, Z.~H.~An$^{1}$, J.~Z.~Bai$^{1}$, Y.~Ban$^{18}$,
N.~Berger$^{1}$, J.~M.~Bian$^{1}$, I.~Boyko$^{13}$
% ...
\\
\vspace{0.2cm}
(BESIII Collaboration)\\
\vspace{0.2cm}
{\it
$^{1}$ Institute of High Energy Physics, Beijing 100049, P. R. China\\
% ...
$^{5}$ G.I. Budker Institute of Nuclear Physics SB RAS (BINP), Novosibirsk 630090, Russia\\
% ...
$^{9}$ Guangxi University, Naning 530004, P. R. China\\
% ...
$^{13}$ Joint Institute for Nuclear Research, 141980 Dubna, Russia\\
% ...
$^{18}$ Peking University, Beijing 100871, P. R. China\\
% ...
$^{31}$ University of Science and Technology of China, Hefei 230026, P. R. China\\
% ...
}}
\vspace{0.4cm}
}


\begin{abstract}
We present measurements of the charmonium state $h_c(^1P_1)$ made
with 106M $\psi'$ events collected by BESIII at BEPCII. Clear
signals are observed for $\psi'\to\pi^0 h_c$ with and without the
subsequent radiative decay $h_c\to\gamma\eta_c$. First measurements
of the absolute branching ratios $\mathcal{B}(\psi' \rightarrow
\pi^0 h_c) = (8.4 \pm 1.3 \pm 1.0) \times 10^{-4}$ and
$\mathcal{B}(h_c \rightarrow \gamma \eta_c) = (54.3 \pm 6.7 \pm
5.2)\%$ are presented.  A statistics-limited determination of the
previously unmeasured $h_c$ width leads to an upper limit
$\Gamma(h_c)<1.44$~MeV (90\% confidence).  Measurements of $M(h_c) =
3525.40 \pm 0.13 \pm 0.18$~MeV/$c^2$ and $\mathcal{B}(\psi'
\rightarrow \pi^0 h_c) \times \mathcal{B}(h_c \rightarrow \gamma
\eta_c) = (4.58 \pm 0.40 \pm 0.50) \times 10^{-4}$ are consistent
with previous results.

\end{abstract}


\maketitle

Although the charmonium family of mesons composed of a charmed quark
and its own antiquark~($c\bar{c}$) has been studied for many years,
knowledge is sparse on the singlet state $h_c(^1P_1)$. The only known production
mode of $h_c$ from other charmonium decays is $\psi'\to \pi^0 h_c$,
but its branching ratio has not been previously measured. 
For the decay chain $\psi' \to \pi^0 h_c$, $h_c\to\gamma\eta_c$, the absolute
branching ratio of $h_c\to\gamma\eta_c$ also has not previously been
measured.  Their measurements will allow the test of isospin
violation mechanisms in charmonium hadronic transitions and guide
refinements of theoretical methods in the charmonium region. 
Early predictions for the properties of the $h_c$ are found in
Refs.~\cite{KYT,Ko}.  More recently, Kuang
~\cite{ref:hctheorykuang02} considered the effect of $S-D$ mixing
and predicted
$\mathcal{B}(\psi'\to\pi^{0}h_c)=(0.4-1.3)\times10^{-3}$, and gave
estimates of $\mathcal{B}(h_c\to\gamma\eta_c)=88\%$ and
$\Gamma(h_c)=(0.51 \pm 0.01)$ MeV for perturbative QCD (PQCD)
and $\mathcal{B}(h_c\to\gamma\eta_c)=41\%$ and $\Gamma(h_c)=(1.1 \pm
0.09)$ MeV with nonrelativistic QCD (NRQCD).  Godfrey and
Rosner have predicted
$\mathcal{B}(h_c\to\gamma\eta_c)=38\%$~\cite{ref:hctheorygod02}. A
recent unquenched lattice QCD analysis~\cite{Dudek:2006ej} included
a prediction of the width
$\Gamma(h_c\to\gamma\eta_c)=(0.601\pm0.055)$~MeV.

Information about the spin-dependent interaction of heavy quarks
can be obtained from precise measurement of the $1P$ hyperfine mass
splitting $\Delta~M_{hf}\equiv\langle M(1^3P)\rangle- M(1^1P_1)$,
where $\langle
M(1^3P_{J})\rangle=(M(\chi_{c0})+3M(\chi_{c1})+5M(\chi_{c2}))/9=
3525.30\pm0.04$~MeV/$c^2$~\cite{ref:pdg} is the spin-weighted centroid of
the $^3P_J$ mass and $M(1^1P_1)$ is the mass of the singlet state
$h_c$. A non-zero hyperfine splitting may give indication of nonvanishing
spin-spin interactions in charmonium potential models~\cite{swanson}.

This Letter reports first results from the BESIII experiment at the
BEPCII storage ring~\cite{ref:bes3,ref:bes3physics} on the
production and decay of the $h_c$ at the $\psi'$ resonance.  We study
distributions of mass recoiling against a detected $\pi^0$ to measure
$\psi'\to\pi^0 h_c$ both inclusively and in events tagged as
$h_c\to\gamma\eta_c$ by detection of the $E1$ transition photon.
Combining inclusive and $E1$-tagged yields, we determine for the first
time the branching ratio for $\psi'\to\pi^0 h_c$ and that for the $E1$ 
transition $h_c\to\gamma\eta_c$, as well as the
$h_c$ width.  We also measure the product branching ratio for the chain 
$\psi'\to\pi^{0}h_c$, $h_c\to\gamma\eta_c$ and the $h_c$ mass, confirming
previous results.

The CLEO Collaboration first observed the $h_c$ in the cascade process $\psi'\to\pi^{0}h_c$,
$h_c\to\gamma\eta_c$ in both inclusive and exclusive measurements~\cite{ref:cleohc05}, 
and later improved the $h_c$ mass determination~\cite{ref:cleohc08} with more
data.  They average their measurements
in~\cite{ref:cleohc08} to obtain $M(h_c) = (3525.20 \pm 0.18 \pm
0.12)$~MeV/$c^2$. The E835 experiment~\cite{ref:E835hc} scanned
antiproton energy and observed $p\bar{p} \to h_c \to
\gamma\eta_c$.  Recently, CLEO reported evidence
for the decay $h_c \to \pi^+\pi^-\pi^+\pi^-\pi^0$ with indications
that the width for $h_c$ multihadronic decays is comparable to that
for the radiative transition to $\eta_c$~\cite{ref:cleohc09}.

BEPCII is a two-ring $e^+e^-$ collider designed for a peak 
luminosity of $10^{33}$ cm$^{-2}s^{-1}$ at a beam current of 0.93~A.
The cylindrical core of the BESIII detector consists of a helium-gas-based drift chamber~(MDC), a
plastic scintillator Time-of-Flight system~(TOF), and a CsI(Tl) Electromagnetic
Calorimeter~(EMC), all enclosed in a superconducting solenoidal magnet providing
a 1.0-T magnetic field.  The solenoid is supported by an octagonal  flux-return yoke
with resistive plate counter muon identifier modules (MU) interleaved with steel.
The charged particle and photon acceptance is $93\%$ of $4\pi$, and
the charged particle momentum and photon energy resolutions at 1~GeV are $0.5\%$
and $2.5\%$, respectively.


We perform the analysis on a data sample consisting of $(1.06 \pm 0.04) \times 10^8$ $\psi'$ decays~\cite{ref:chicpaper}. An independent
sample of $42.6~\rm{pb}^{-1}$ at 3.65~GeV is used to determine
continuum ($e^{+}e^{-}\to q\bar{q}$) background. We measure $h_c$
production by selecting events consistent with $\psi'\to\pi^0h_c$
(momentum $p(\pi^0)\simeq84$~MeV/$c$) and fitting the distribution
of masses recoiling against the $\pi^0$. The yield of $\psi'\to
\pi^0h_c, h_c\to\gamma\eta_c$ is determined with the same technique
on events containing a $\sim 500$~MeV photon.

We model BESIII with a Monte Carlo (MC) simulation based on
Geant4~\cite{Agostinelli:2002hh,Allison:2006ve}.
EvtGen~\cite{ref:bes3gen} is used to generate $\psi'\to\pi^{0}h_c$
events with an $h_c$ mass of 3525.28~MeV/$c^2$~\cite{ref:cleohc08} and a
width equal to that of the $\chi_{c1}$ (0.9~MeV).  The $E1$
transition $h_c\to \gamma\eta_c$ (assumed branching ratio $50\%$) is
modeled with EvtGen, with an angular distribution in the $h_c$ frame
of $1+\cos^2\theta$.  Other $h_c$ decays are simulated by
PYTHIA~\cite{ref:bes3gen}.  The $\eta_c$ decay parameters are set to
Particle Data Group values~\cite{ref:pdg}, with known modes
simulated by EvtGen and the remainder by PYTHIA.  Backgrounds are
studied with a sample of $\psi'$ generated by KKMC~\cite{ref:kkmc}
with known decays modeled by EvtGen and other modes generated with
Lundcharm~\cite{ref:bes3gen}.

Charged tracks in BESIII are reconstructed from MDC hits.  To
optimize the momentum measurement, we select tracks in the polar
angle range $|\cos\theta|~<~0.93$ and require that they pass within
$\pm 10$~cm of the interaction point in the beam direction and
within $\pm 1$~cm in the plane perpendicular to the beam.
Electromagnetic showers are reconstructed by clustering EMC crystal
energies.  Efficiency and energy resolution are improved by
including energy deposits in nearby TOF counters.  Showers used in
selecting $E1$-transition photons and in $\pi^0$ reconstruction must
satisfy fiducial and shower-quality requirements.  Showers in the
barrel region ($|\cos\theta|<0.8$) must have a minimum energy of
25~MeV, while those in the endcaps ($0.86 < |\cos\theta| < 0.92$)
must have at least 50~MeV.  Showers in the region between the
barrel and endcap are poorly reconstructed and are excluded. To 
eliminate showers from charged particles, a photon must
be separated by at least 10$^\circ$ from any charged track.  EMC
cluster timing requirements suppress electronic noise and energy
deposits unrelated to the event. Diphoton pairs are accepted as
$\pi^0$ candidates if their reconstructed mass satisfies $120 <
M_{\gamma \gamma} < 145$~MeV/$c^2$, approximately equivalent to 1.5
(2.0) standard deviations on the low-mass (high-mass) side of the
mass distribution. A 1-C kinematic fit with the $\pi^0$ mass
constrained to its nominal value is used to improve the energy
resolution.

Candidate events must have at least two charged tracks, with at
least one passing the fiducial and vertex cuts.  For selection of
inclusive $\pi^0$ events we demand at least two photons passing the
above requirements, with at least three photons for $E1$-tagged
candidate events.  To suppress continuum background, the total
energy deposition in the EMC must be greater than 0.6~GeV.  Background
events from  $\psi'\to\pi^+\pi^- J/\psi$ and $\pi^0\pi^0 J/\psi$ are
suppressed by requiring that the $\pi^+\pi^-$ ($\pi^0\pi^0$) recoil
mass be outside the range $3097 \pm 7$~MeV/$c^2$ ($3097 \pm 15$~MeV/$c^2$).

To improve the signal-to-noise ratio, photons used in signal $\pi^0$
candidates must be in the barrel and have energies
greater than 40~MeV.  For the inclusive analysis, $\pi^0$ candidates
are excluded if either daughter photon can make a $\pi^0$ with
another photon in the event. Figure~\ref{hc-FitData} shows the
inclusive $\pi^0$ recoil mass spectra after applying the above
selection criteria. For the $E1$-tagged selection
(Fig.~\ref{hc-FitData} (a)), we require one photon in the energy
range $465-535$~MeV, demanding that it not form a $\pi^0$
with any other photon in the event. Because $E1$-tagged events
have reduced background, we keep them even if daughter photons can be 
used in more than one $\pi^0$ combination, choosing the candidate 
with the minimum 1-C fit $\chi^2$. Events with more than one $\pi^0$ in
the $3.500-3.555$~GeV/$c^2$ recoil-mass region are excluded.

\begin{figure}[htbp]
\begin{center}
  \includegraphics[width=8cm]{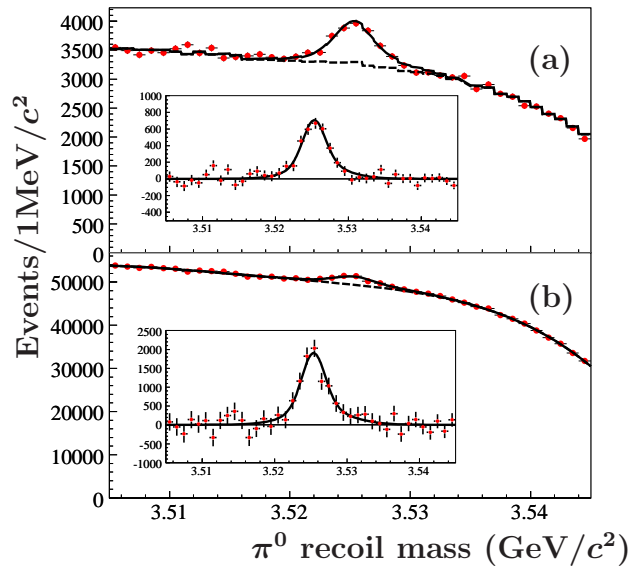} \put(-40,180){\bf
    \large~(a)} \put(-40,90){\bf
    \large~(b)}\put(-230,70){\rotatebox{90}{\bf \boldmath \large
      Events/1MeV/$c^2$}}\put(-140,-5){\bf\boldmath \large $\pi^{0}$
    recoil mass~(GeV/$c^2$)}
\caption{(a)~The $\pi^0$ recoil mass spectrum and fit for the
  $E1$-tagged analysis of $\psi'\to \pi^0h_c,
  h_c\to\gamma\eta_c$;~(b)~the $\pi^0$ recoil mass spectrum and fit for
  the inclusive analysis of $\psi'\to\pi^{0}h_c$. Fits are shown as
  solid lines, background as dashed lines. The insets show the 
  background-subtracted spectra.}\label{hc-FitData}
\end{center}
\end{figure}

The $\pi^0$ recoil mass spectra (Fig.~\ref{hc-FitData}) are fitted
by an unbinned maximum likelihood method.  Because of its lower
background, the $E1$-tagged fit is used to extract the mass and
width of the $h_c$, which are then fixed for the inclusive fit.  For
the $E1$-tagged fit, the signal is parameterized as a Breit-Wigner
function with the mass and width free, convoluted with a detector
resolution function obtained from MC simulation. The background shape is 
obtained from the $\pi^0$ recoil mass spectrum with no photons in the 
signal region of $400 - 600$~MeV and at least one good 
photon in the signal-free region below $400$~MeV and above $600$~MeV. The upper and lower limits of the
accepted ranges were varied to assess possible systematic
uncertainty. The results of this fit are a yield of $E1$-tagged
$h_c$ decays of $N^{E1}=3679\pm319$ and $h_c$ parameters
$M(h_c)=3525.40\pm0.13$~MeV/$c^2$ and
$\Gamma(h_c)=0.73\pm0.45$~MeV, where the errors are
statistical. The fit quality assessed with the binned distribution
of Fig.~\ref{hc-FitData}(a) is $\chi^2/d.o.f.=33.5/36$~($p$-value $58.8\%$), and the statistical
significance of the $h_c$ signal is 18.6$\sigma$.  The fit of the
inclusive $\pi^0$ spectrum in Fig.~\ref{hc-FitData}~(b) is performed
similarly, except that the $h_c$ mass and width are fixed and the
background is described by a 4th-order Chebychev polynomial with all
parameters free. The fit result for the inclusive $h_c$ yield is
$N^{inc}=10353 \pm 1097$, with $\chi^2/d.o.f.=24.5/34$~($p$-value
$88.4\%$) and $9.5\sigma$ statistical significance. The insets of
Fig.~\ref{hc-FitData} show the $\pi^0$ recoil-mass spectra with the
fitted backgrounds subtracted.

The product branching ratio $\mathcal{B}_{1}(\psi'\to\pi^0h_c)\times
\mathcal{B}_{2}(h_c\to\gamma\eta_c)$ depends on the number of
$\psi'$ decays in the sample and the yield and detection efficiency
for $E1$-tagged events~($\epsilon_{12}$), as given by
Eq.~(\ref{B1B2}):
\begin{equation}\label{B1B2}
\displaystyle \mathcal{B}_{1}\times\mathcal{B}_{2}=
\frac{N^{E1}}{\epsilon_{12}\times N(\psi')}.
\end{equation}
\noindent The efficiency, determined with the signal MC, is
$\epsilon_{12}=7.57\%$. The branching ratios for the inclusive
process $\mathcal{B}_{1}(\psi'\to\pi^0h_c)$ and for the $E1$
transition $\mathcal{B}_{2}(h_c\to\gamma\eta_c)$ are related to the
inclusive yield $N^{inc}$ and the efficiencies for selecting $h_c$
decays to $\gamma\eta_c$~($\epsilon_{1}^{E1}$) and to other final
states~($\epsilon_{1}^{had}$), as given by Eq.~(\ref{B1B22}):
\begin{equation}\label{B1B22}
\displaystyle\mathcal{B}_{1}=\frac{N^{inc}}{(\epsilon_{1}^{E1}\mathcal{B}_{2}+\epsilon_{1}^{had}(1-\mathcal{B}_{2}))\times
N(\psi')}.
\end{equation}
\noindent The detection efficiencies are $\epsilon_{1}^{E1}=12.89\%$
and $\epsilon_{1}^{had}=10.02\%$, respectively.

Using the numbers obtained above, we find
$\mathcal{B}_{1}=(8.4\pm1.3)\times10^{-4}$,
$\mathcal{B}_{2}=(54.3\pm6.7)\%$, and $\mathcal{B}_{1}\times
\mathcal{B}_{2}=(4.58\pm0.40)\times10^{-4}$, where the errors are
statistical only.

Systematic uncertainties for our measurements are summarized in
Table~\ref{tab-syshc}. Dominant sources are the treatment of the
background in the recoil-mass fits and imperfect modeling of
photon and $\pi^0$ detection in BESIII.

\begin{table*}[htbp]
\small
\begin{center}
\caption{Summary of systematic errors.} \label{tab-syshc}
\vspace{0.2cm}
\begin{tabular}{cccccc}
\hline \hline

Source & $M(h_c)$(MeV/$c^2$) & $\Gamma(h_c)$(MeV) &
$\mathcal{B}_{1}$($10^{-4}$) & $\mathcal{B}_{1}\times\mathcal{B}_{2}$($10^{-4}$) & $\mathcal{B}_{2}$($\%$)\\
\hline
\hline Background shape and fit range  & 0.11 & 0.23 & 0.4 & 0.22 & 4.4\\
Energy scale, position reconstruction and 1-C fit  & 0.13 & $0.06$ & 0.5 & 0.10 & 2.1 \\
Energy resolution & 0.00 & $0.15$  & 0.2 & 0.03 & 1.0 \\
Background veto & 0.05 & 0.03 & 0.0 & 0.03 & 0.3\\
$\pi^0$ efficiency & 0.00 & 0.00 & 0.3 & 0.14 & 0.0\\
 $E1$ photon efficiency & 0.00 & 0.00 & 0.0 & 0.10 & 1.2\\
Number of $\pi^0$ & 0.00 & 0.00 & 0.6 & 0.35 & 0.6\\
Number of charged  tracks & 0.00 & 0.00 & 0.1 & 0.06 & 0.1\\
 $N(\psi')$ & 0.00 & 0.00 & 0.4 & 0.19 & 0.0\\
 $M(\psi')$ & 0.03 & 0.02 & 0.0 & 0.00 & 0.0\\
 $M(\eta_c)$ and $\Gamma(\eta_c)$ & 0.00 & 0.00 & 0.0 & 0.01 & 0.3\\
\hline
Total systematic error & 0.18 & 0.28 & 1.0 & 0.50 & 5.2\\
\hline
\end{tabular}
\end{center}
\end{table*}



For the inclusive measurements, we explore sensitivity to the
background parameterization by changing the order of the Chebychev
polynomial from 4 to 5 and by considering alternative fitting
functions based on MC simulations. For the $E1$-tagged measurements,
alternative background shapes are obtained by varying the
photon-energy boundaries defining the signal-free sample. Systematic
uncertainties are set based on the largest changes observed in the
measured quantities for all alternative backgrounds. The uncertainty
due to the choice of the fitting range is evaluated by changing from
3505$-$3545~MeV/$c^2$ to 3500$-$3540~MeV/$c^2$ and
3510$-$3545~MeV/$c^2$.

Our analysis depends on accurate simulation of the detector response 
for shower energy measurements. The calibration uncertainty in the  
photon-energy scale is estimated to be $\pm 0.4\%$ by studying 
$\psi'\to\gamma\chi_{c1,2}$  and radiative Bhabha events. 
Studies of the energy spectra for photons in radiative 
$\psi'$ decays show the energy resolution to be larger in data than 
MC by $4\%$ for $\psi'\to\gamma\chi_{c1}$ and $2\%$ for
$\psi'\to\gamma\chi_{c2}$. We estimate systematic uncertainties due
to the energy measurement by determining the changes in results
after adjusting the photon response accordingly.  We also did more
extensive studies allowing for correlations among the different
effects by simultaneously varying the energy scale, energy
resolution, reconstructed position, and error matrix of the photon
measurement.  These studies gave a somewhat larger uncertainty for
the $h_c$ mass. The maximum observed change in the $h_c$ mass is
0.13~MeV/$c^2$, which we take as its systematic uncertainty due to
the energy measurement.

We estimate the uncertainty in simulating the $E1$-photon selection 
efficiency with $e^+e^-\to\gamma e^+e^-$ events, studying the ratio 
$E_{meas}$/$E_{exp}$ of measured to expected photon energy, where 
$E_{exp}$ is determined from the $e^+e^-$ recoil energy.  Comparing 
this ratio between data and MC provides a smearing function that is 
used as an alternative to the standard line shape. This modification 
results in a $2\%$ change in the efficiency for $E1$-photon selection, and
associated systematic uncertainties are obtained by varying
$\epsilon_{12}$ by $\pm 2\%$.

The photon detection efficiency and resolution also enter through the
uncertainty in the reconstruction efficiency of the $\pi^0$ selection,
which was determined to be $\pm 3\%$ by analyzing $\psi'\to\pi^0\pi^0
J/\psi, J/\psi\to l^+l^-$ in data and MC. Systematic errors are obtained by varying the efficiencies
$\epsilon_{1}^{E1}$, $\epsilon_{1}^{had}$ and $\epsilon_{12}$
simultaneously by $\pm 3\%$.  The efficiency uncertainty due to the
simulation of the number of $\pi^0$s, which is mainly generator
dependent, is estimated by a comparison between data and MC for
$\psi'$ decays, which we assume behave similarly to $h_c$ decays.
Variations in the efficiencies $\epsilon_1^{E1}$, $\epsilon_1^{had}$
and $\epsilon_{12}$ are determined by the equation $\displaystyle
\Delta\epsilon= \sum\epsilon_{i}\times\Delta N^{\pi^0}_{i}$, where
$\Delta\epsilon$ denotes the difference between the efficiencies from
data and MC simulations, $\epsilon_{i}$ is the efficiency when
$N_{\pi^0}=i$ in the event, and $\Delta N^{\pi^0}_{i}$ is the relative
difference for $N_{\pi^0}=i$.  The systematic errors are obtained by
simultaneously varying $\epsilon_{1}^{E1}$, $\epsilon_{1}^{had}$ and
$\epsilon_{12}$ by $\Delta\epsilon_{1}^{E1}$,
$\Delta\epsilon_{1}^{had}$ and $\Delta\epsilon_{12}$.

Other sources of systematic uncertainties are found to be small. The
uncertainty in the efficiency of the requirement on the number of
charged tracks arises from uncertainty in simulating $h_c$
decays and in modeling charged-particle detection.  We find that
$9\%$ of simulated $h_c\to\gamma\eta_c$ events and $5.5\%$ of
other $h_c$ decays fail the requirement on the number of charged
tracks. For generic $\psi'$ decays we find relative differences
between data and MC in the corresponding efficiencies to be less
than $10\%$. Assuming similar consistency for $h_c$ decays, we
simultaneously vary $\epsilon_{12}$ and $\epsilon_1^{E1}$ by
$9\%\times10\%$=$0.9\%$, and $\epsilon_1^{had}$ by
$5.5\%\times10\%=0.55\%$ to estimate the resulting systematic
uncertainty in the branching ratios. Systematic uncertainties
associated with the requirements to suppress $\psi'$ to $J/\psi$
hadronic transitions are shown to be negligible for all measurements
by varying the excluded recoil-mass range. The $\pm 4\%$
uncertainty in the number of $\psi'$ in our sample makes a small
contribution to the overall uncertainty for the measured branching
ratios.  Uncertainty in the $\psi'$ mass has negligible effect.
Assumptions for the $\eta_c$ mass and width in signal simulations
affect detection efficiencies through the $E1$-photon energy.
Associated systematic uncertainties are set by varying these
parameters within errors, recalculating efficiencies, and
determining the maximum changes in the branching ratios.

We treat all sources of systematic uncertainty as uncorrelated and
combine in quadrature to obtain the overall systematic uncertainties
and the following results: $\displaystyle
M(h_c)=3525.40\pm0.13\pm0.18$~MeV/$c^2$, $\displaystyle
\Gamma(h_c)=0.73\pm0.45\pm0.28$~MeV ($<1.44$~MeV~at 90\%
confidence),
$\mathcal{B}(\psi'\to\pi^{0}h_c)=(8.4\pm1.3\pm1.0)\times10^{-4},$
$\mathcal{B}(\psi'\to\pi^{0}h_c)\times\mathcal{B}(h_c\to\gamma\eta_c)=(4.58\pm0.40\pm0.50)\times10^{-4}$,
and $\displaystyle
\mathcal{B}(h_c\to\gamma\eta_c)=(54.3\pm6.7\pm5.2)\%.$ In all cases
the first errors are statistical and the second systematic.
Our measurements of  $\mathcal{B}(\psi'\to\pi^{0}h_c)$ and 
$\mathcal{B}(h_c\to\gamma\eta_c)$ and information about the 
$h_c$ width are the first experimental results for these quantities. 
The determinations of $M(h_c)$ and 
$\mathcal{B}(\psi'\to\pi^{0}h_c)\times\mathcal{B}(h_c\to\gamma\eta_c)$
are consistent with published CLEO results~\cite{ref:cleohc08} and of comparable precision.

Comparing our results for $h_c\to\gamma\eta_c$ to the $E1$ radiative
transitions $\chi_{c1}\to{\gamma}J/\psi$, we find that the branching ratio
$\mathcal{B}(h_c\to\gamma\eta_c)$ is consistent with the PDG value for 
$\mathcal{B}(\chi_{c1}\to{\gamma}J/\psi)=(36.0\pm1.9)\%$~\cite{ref:pdg}; the total widths
$\Gamma(\chi_{c1})$ and $\Gamma(h_{c})$ are also consistent. Our result
for $\mathcal{B}(h_c\to\gamma\eta_c)$ is close to the prediction of
Ref.~\cite{ref:hctheorygod02} ($38\%$) and the NRQCD prediction of
Ref.~\cite{ref:hctheorykuang02} (41\%).  The branching ratio
$\mathcal{B}(\psi'\to\pi^{0}h_c)$ is consistent with the prediction of
Ref.~\cite{ref:hctheorykuang02} $((0.4-1.3)\times10^{-3})$, and the
total width $\Gamma(h_c)$ is consistent with the predictions of 
Refs.~\cite{ref:hctheorykuang02} and \cite{Dudek:2006ej}.  We find the 
$1P$ hyperfine mass splitting to be $\displaystyle \Delta~M_{hf} \equiv
\langle M(1^3P) \rangle - M(1^1P_1) =-0.10 \pm 0.13
\pm0.18$~MeV/$c^2$, consistent with no strong spin-spin interaction.

We thank the accelerator group and computer staff of IHEP for
extraordinary effort in producing beams and processing data.  We are
grateful for support from our institutes and universities and from
these agencies: Ministry of Science and Technology of China, National
Natural Science Foundation of China, Chinese Academy of Sciences,
Istituto Nazionale di Fisica Nucleare, Russian Foundation for Basic
Research, Russian Academy of Science (Siberian branch), U. S.
Department of Energy, and National Research Foundation of Korea.



\begin{thebibliography}{99}
%
\bibitem{KYT} Y.~P. Kuang, S.~F. Tuan, and T.~M. Yan, Phys. Rev. D
  {\bf 37}, 1210 (1988).
\bibitem{Ko} P. Ko, Phys. Rev. D {\bf 52}, 1710 (1995).
\bibitem{ref:hctheorykuang02} Y.~P.~Kuang,~Phys.\ Rev.\ D\ {\bf 65}, 094024~(2002).
\bibitem{ref:hctheorygod02} S.~Godfrey and J.~Rosner, Phys.\ Rev.\ D\ {\bf 66},~014012~(2002).
\bibitem{Dudek:2006ej}
  J.~J.~Dudek, R.~G.~Edwards and D.~G.~Richards,
  Phys.\ Rev.\  D {\bf 73}, 074507 (2006).

\bibitem{ref:pdg} C.~Amsler~{\it et al.} (Particle Data Group), Phys.\ Lett.\ B\ {\bf 667}, 1~(2008).
\bibitem{swanson} See, for example, E. S. Swanson, Phys. Rep. {\bf
    429}, 243 (2006), and references therein.
    
\bibitem{ref:bes3} M.~Ablikim~{\it et al.} (BESIII Collaboration),~ Nucl.\ Instrum.\ Meth.\ A. {\bf 614}, 3~(2010).    
\bibitem{ref:bes3physics} ``Physics at
BESIII'', Edited by K.~T.~Chao and Y.~F.~Wang, Int.~J.~Mod.\ Phys.\
A\ {\bf 24}, No.1(2009) supp. 

\bibitem{ref:cleohc05} J.~L.~Rosner~{\it et al.} (CLEO Collaboration), Phys.\ Rev.\ Lett.\ {\bf 95}, 102003 (2005); P.~Rubin {\it et
al.} (CLEO Collaboration), Phys. Rev. D {\bf 72}, 092004 (2005).
\bibitem{ref:cleohc08} S.~Dobbs~{\it et al.} (CLEO Collaboration), Phys.\ Rev.\ Lett.\ {\bf101}, 182003~(2008).
\bibitem{ref:E835hc}M.~Andreotti~{\it et al.} (E-835 Collaboration), Phys.\ Rev.\ D\ {\bf72}, 032001~(2005).
\bibitem{ref:cleohc09} G.~S.~Adams~{\it et al.} (CLEO Collaboration), Phys.\ Rev.\ D\ {\bf 80},~051106~(2009).

\bibitem{ref:chicpaper} M.~Ablikim~{\it et al.} (BESIII
  Collaboration), submitted to Phys. Rev. D, arXiv:1001.5360.

\bibitem{Agostinelli:2002hh}
  S.~Agostinelli {\it et al.} (GEANT4 Collaboration),
  Nucl.\ Instrum.\ Meth.\  A~{\bf 506}, 250 (2003).
\bibitem{Allison:2006ve}
  J.~Allison {\it et al.},
  IEEE Trans.\ Nucl.\ Sci.\ {\bf 53}, 270 (2006).
\bibitem{ref:bes3gen}R. G. Ping, Chinese Physics C {\bf 32}, 8~(2008).
\bibitem{ref:kkmc} S.~Jadach, B.~F.~L.~Ward and Z.~Was,  Comput.\ Phys.\ Commun.\  {\bf 130}, 260
(2000); S.~Jadach, B.~F.~L.~Ward and Z.~Was, Phys.\ Rev.\  D {\bf
63}, 113009 (2001).


\end{thebibliography}
\end{document}